\documentclass[oribbl]{llncs}

\usepackage[dvipsnames]{xcolor}
\RequirePackage{filecontents}

\usepackage[
	backend=biber,
	style=numeric,
	firstinits=true,
	url=false,
	isbn=true,
	]{biblatex}

\newcommand{\theReferenceText}{\fullcite{GreinerPetter2018}}
\usepackage{comgipp}

\usepackage[dvipsnames]{xcolor}

\usepackage{lstomdoc}
\usepackage{enumitem}
\usepackage[subtle]{savetrees}

\setlist{leftmargin=5mm}

\usepackage{graphicx}
\usepackage{hyperref}
\usepackage{latexml}

\newcommand{\name}{MathTools}
\newcommand{\element}[1]{\textbf{#1}}

\newcommand{\moduleSec}[1]{{\noindent\bf #1:}}
\begin{document}
\renewcommand{\thelstlisting}{\arabic{lstlisting}} % change listing numbering to global numbering

\title{\textit{\name}: An Open API for Convenient \MathML{} Handling}

% Authors are joined by \and. Their affiliations are given by \inst, which indexes
% into the list defined using \institute
%
\author{
   Andr\'{e} Greiner-Petter\,\inst{1}
\and
   Moritz Schubotz\,\inst{1}
\and
   Howard S.~Cohl\,\inst{2}
\and
   Bela Gipp\,\inst{1}
}

% Institutes for affiliations are also joined by \and,
\institute{
	\hspace{-0.15cm}
	Dept.~of Computer and Information Science,\\
	University of Konstanz, Box 76, 78464 Konstanz, Germany,\\
	\email{\{first.last\}@uni-konstanz.de}
\and
	%\hspace{-0.10cm}
	Applied and Computational Mathematics Division,
    National Institute of Standards and Technology, Mission Viejo, CA 92694, U.S.A.,
	\email{howard.cohl@nist.gov}
}
\authorrunning{Greiner-Petter, Schubotz, Cohl, and Gipp}

\clearpage

%%%%%%%%%%%%%%%%%%%%%%%%%%%%%%%%%%%%%%%%%%%%%%%%%%%
\maketitle
\thispagestyle{firststyle}
%%%%%%%%%%%%%%%%%%%%%%%%%%%%%%%%%%%%%%%%%%%%%%%%%%%

%\vspace{-0.665cm}
\begin{abstract}
Mathematical formulae carry complex and essential semantic information in a variety of formats.
Accessing this information with different systems requires a standardized machine-readable format that is capable of encoding presentational and semantic information.
Even though \MathML{} is an official recommendation by W3C and an ISO standard 
for representing mathematical expressions, we could identify only very few systems which use the full descriptiveness of \MathML.
\MathML's high complexity results in a steep learning curve for novice users.
We hypothesize that this complexity is the reason why many community-driven projects refrain from using \MathML, and instead develop problem-specific data formats for their purposes.
We provide a user-friendly, open-source application programming interface for controlling \MathML{} data.
Our API is written in JAVA and allows to create, manipulate, and efficiently access commonly needed information in presentation and content \MathML.
Our interface also provides tools for calculating differences and similarities between \MathML{} expressions.
The API also allows to determine the distance between expressions using different similarity measures.
In addition, we provide adapters for numerous conversion tools and the canonicalization project.
Our toolkit facilitates processing of mathematics for digital libraries, without the need to obtain XML expertise.
\end{abstract}

%\vspace{-0.7cm}
%\vspace{-0.1cm}
%------------------------------------------------------------------------------

%\vspace*{-1em}
\section{Introduction}
%\vspace*{-0.7em}
\MathML{} has the ability to represent presentational and content information.
Several optional features and markup options make \MathML{} a highly versatile, but complex format.
\MathML's ability to mix presentation and content markup for an expression (hereafter called fully descreptive \MathML{}) makes \MathML{} increasingly important for mathematical digital libraries.
For example, Listing~\ref{lst.MathML} shows a simple example of a cross-referenced parallel markup \MathML{} document generated from the \LaTeX{} string \verb|\frac{a}{b}|.

\begin{lstlisting}[label={lst.MathML},mathescape=true,float,caption=Parallel markup MathML with examples of cross-references.]
<math><semantics>
  <mfrac id="p.2" xref="c.1">
  <mi id="p.1" xref="c.2">a</mi>
  <mi id="p.3" xref="c.3">b</mi></mfrac>
<annotation-xml encoding="MathML-Content"><apply>
  <divide id="c.1" xref="p.2"/>
  <ci id="c.2" xref="p.1">a</ci>
  <ci id="c.3" xref="p.3">b</ci></apply></annotation-xml>
<annotation encoding="application/x-tex">\frac{a}{b}</annotation>
</semantics></math>
\end{lstlisting}

Although \MathML{} is an official recommendation of the \href{https://www.w3.org/Math/}{World Wide Web Consortium} since 1998, has been an ISO standard (ISO/IEC 40314) since 2015, and is also part of HTML5, it is still a rarely used format.
For example, the prominent browser Microsoft Internet Explorer\footnote{The mention of specific products, trademarks, or brand names is for purposes of identification only. Such mention is not to be interpreted in any way as an endorsement or certification of such products or brands by the National Institute of Standards and Technology, nor does it imply that the products so identified are necessarily the best available for the purpose. All trademarks mentioned herein belong to their respective owners.} does not support \MathML.
Google Chrome supported \MathML{} only in version 24 and dropped it again in newer versions in favor of MathJax\footnote{\url{https://bugs.chromium.org/p/chromium/issues/detail?id=152430\#c43}}.
Furthermore, we were only able to identify few databases
%m:I am not entirely sure if Ginev uses cmml. I guess not
% (such as arXMLiv~\cite{DBLP:conf/gi/GinevJAGDK09})
that use fully descriptive \MathML{}.
Most databases use basic \MathML{} instead, such as the DLMF~\cite{DLMF} that only provides presentational \MathML{}.
Numerous tools that process math allow \MathML{} as an input or export format.
Most of these tools avoid \MathML{} as an internal data representation.
Instead, many systems create their own problem-specific data representation and implement custom tools to parse \MathML{} for internal processing.
%We hypothesize that the complexity of \MathML{} and the steep learning curve required to master \MathML{} are the main reasons why community-driven projects usually avoid the use of content \MathML{} in their mathematical databases.
This workaround causes problems in regard to reusability, restricts applicability, and decreases efficiency for processing mathematics.

During our research of a \MathML{} benchmark~\cite{Schubotz2018}, we realized that parsing \MathML{} data is an error-prone endeavor.
Even small changes, such as missing default namespaces, can cause the parsing process to fail.

With \name{}, we provide a collection of tools that overcomes the issues of complexity and simplifies the access to \MathML{} data.
\name{} also contains an open API for easy access to useful services that we used in several of our previous projects, such as for similarity calculations or for \LaTeX{} to \MathML{} conversion 
tools~\cite{DBLP:conf/mkm/CohlSMSZMD15,
DBLP:conf/cikm/MeuschkeSHSG17,
Meuschke2018a,
disCicm14Mlp,
DBLP:phd/dnb/Schubotz17,
DBLP:conf/clef/SchubotzKMHG17,
Schubotz2018,
disSigir16,
vmext17}.
The tools we have developed are able to parse fully descriptive \MathML{} and to a certain degree, invalid \MathML{}.
Furthermore, we provide easy access to several useful features implemented in \MathML{} related projects.
These features include similarity metrics, distance calculations, and Java imports for state-of-the-art conversion tools.
%\vspace*{-1.5em}
\section{Related Work}\label{sec:rel-work}
%\vspace*{-0.7em}
Many digital libraries do not use \MathML{} but favor their custom internal formats to process mathematics.
The conversion tool \LaTeXML{}~\cite{LaTeXML} is able to generate presentation and content \MathML{} using \LaTeX{} input.
Instead of processing the input directly to \MathML, \LaTeXML{} uses a customized XML format, and creates \MathML{} in a post-processing step.
The Speech Rule Engine (SRE) implemented in MathJax~\cite{Cervone2016} is able to translate presentation \MathML{} expressions to speech.
Translations of the SRE ause a set of rules that are applied to a nonstandard tree representation of the original \MathML{} input.
Computer Algebra Systems (CAS) allow for complex computations of mathematical expressions.
These systems usually create their own internal data format, but typically allow for \MathML{} export.
For example, the CAS MAPLE
parses inputs to a directed acyclic graph (DAG) presentation~\cite[Chapter~2]{MAPLE:ProgrammingGuide}, whereas Mathematica uses a custom tree representation~\cite[Chapter~33]{Wolfram2017}.

{\sloppy
There are other tools that internally use \MathML{} in combination with customized implementations for parsing \MathML{}.
Examples of such tools include \MathML{\sc Can}~\cite{formanek2012}, a canonicalization tool for \MathML{} data;
the Visualization of Mathematical Expression Trees (VMEXT)~\cite{vmext17} for \MathML{} data;
and dependency graphs~\cite{DBLP:journals/ir/KristiantoTA17} for analyzing relationships between math expressions.}
%\vspace*{-1.5em}
\section{MathTools}
%\vspace*{-1.1em} 
We provide \name, a collection of tools for convenient handling of \MathML{} data and an open Java API to access useful services related to \MathML{} data.
\name allows to parse, manipulate, and analyze fully descriptive \MathML{} data conveniently.
We haven chosen Java to realize \name, because the majority of related work is also implemented in Java. Java also is the most frequently used programming language.
The source of this project is publicly available on GitHub\footnote{\url{https://github.com/ag-gipp/MathMLTools}} and on maven-central under an Apache 2 license.

\begin{figure}[!b]
	\center
	\vspace*{-0.7cm}
	\includegraphics[trim=5mm 10mm 5mm 15mm,clip,width=\textwidth]{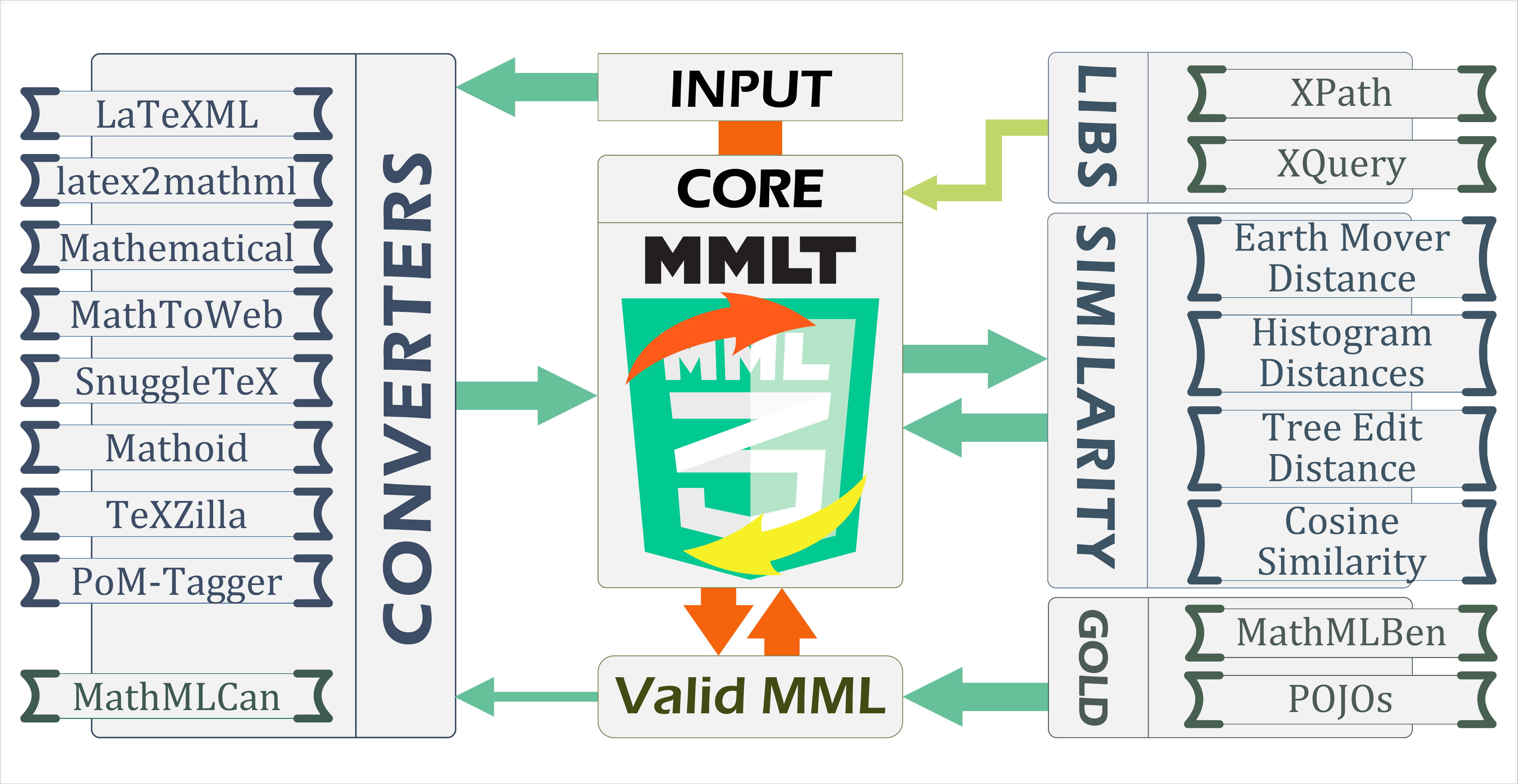}
	\vspace*{-0.7cm}
	\caption{The pipeline schema of \name{} and the modules.
		The orange arrow indicates the general workflow for processing a \MathML{} input to a valid \MathML{} output.
		The converters also allow mathematical \LaTeX{} input, while the \MathML{} feature requires valid \MathML{} input.
		The gold module provides valid \MathML{} without the core module.
		Distances and similarities can be calculated using the similarity module.
		Single elements or subtrees from valid \MathML{} can be accessed through the core module.}
	\label{fig:mmlt} % must be after the caption, otherwise the linked number will be broken
	\vspace*{-0.5cm}
\end{figure}

The project is organized in multiple modules, of which the \element{core} module provides the main functionality.
Figure~\ref{fig:mmlt} illustrates possible workflows for processing the input and shows the interaction of different modules.
The \element{core} module is designed for parsing, loading, storing, and  manipulating presentation and content \MathML{} data.
The \element{core} also provides helper functions for easy access to specific information.
Examples of supported operations include splitting presentation and content trees; accessing original \TeX{} data if available; and extracting all of the identifiers, e.g., \verb|mi| or \verb|ci| elements from the data.
The \element{core} module also allows to `clean' \MathML{} of unwanted features.
Such features include cross-references or entire content subtrees.
This functionality is particularly helpful for obtaining simple \MathML{} for testing and learning purposes.

We equipped \name{} with several extra modules, which we consider useful for typical use cases related to processing \MathML{} data.
\name{} contains the following modules.\\[-0.43cm]

\moduleSec{Gold} is part of the \MathML{\sc Ben} project~\cite{Schubotz2018}, whose purpose is to provide a comfortable interface to \MathML{} gold standards within Java.
The module uses Plain Old Java Objects (POJOs) for representing entries of the gold standard.
The \element{gold} module can be used to implement benchmarks that use \MathML{}.

\moduleSec{Converters} is an outgrowth of the \MathML{\sc Ben} project and provides fast access to state-of-the-art conversion tools.
We implemented a Java interface for each supported tool and the \MathML{\sc Can}~\cite{formanek2012} that allows canocalization of the tool's outputs.
The module allows to:
(1) embed uniform \MathML{} data into translation workflows;
(2) conveniently change the translation engines;
(3) add new conversion tools to the API.

\moduleSec{Libraries} is a collection of reliable XPath and XQuery expressions for accessing elements in \MathML{} data.
Due to the XML structure of \MathML{} documents, \name{} uses data query languages to access specific elements.
The \element{library} module can be used to reliably query data from \MathML{} documents independently of programming languages.

\moduleSec{Similarity} implements several distance and similarity measures for \MathML{} documents.
Comparing \MathML{} data is a common task for many mathematical information retrieval systems, e.g., for mathematical search engines and for plagiarism detection systems that analyze mathematical expressions as well as for many evaluation and benchmark scenarios.
All measures included in the module, except the tree edit distance, produce histograms for \MathML{} elements using the names of the elements and their accumulated frequency of occurrence.
Frequency histograms of \MathML{} elements have been successfully applied as part of plagiarism detection methods to determine the similarity of mathematical expressions, see~\cite{DBLP:conf/cikm/MeuschkeSHSG17,Meuschke2018a} for a more detailed explanation.
We implemented the following similarity measures.
\vspace*{-0.7em}
\begin{itemize}[label={$\bullet$}]
	\item \textbf{Histogram Distance:} calculates the absolute and relative differences of histograms formed from \MathML{} data.
	A small histogram distance indicates semantic similarity between \MathML{} expressions.
	\item \textbf{Tree Edit Distance:} calculates the number of changes that must be performed to transform one \MathML{} tree into another. 
	The user can control the weights for insertions, deletions, and renaming of elements.
	We use an implementation of RTED~\cite{RTED} to perform these calculations. Tree edit distances are helpful
	for detecting structural changes of \MathML{} trees.
	\item \textbf{Earth Mover Distance (EMD):} is originally a distance measure for comparing probability distributions. The measure models a cost function for changing one distribution to another.
	In our case, the probability distribution is the histogram of the \MathML{} data.
	Our calculations are performed using a Java implementation of~\cite{DBLP:conf/iccv/PeleW09}.
	EMD is widely used in multimedia information retrieval and for pattern recognition algorithms.
	\item \textbf{Cosine Similarity:} is a distance measure between two non-zero vectors.
	In our case, the vectors are represented by the histogram of the \MathML{} data.
\end{itemize}
\vspace*{-0.5em}
Note that EMD and Cosine similarity can be used for entire documents.
In this case, the histograms are an accumulation of all \MathML{} expressions in the document.
In this scenario, EMD and Cosine similarities provide a strong measure of how similar documents are in terms of their semantics.
%\vspace*{-1.5em}
\section{Conclusion and Future Work}
%\vspace*{-0.7em} 
The \name{} project is a unified toolkit to facilitate working with \MathML{}.
By using \name{}, other researchers will no longer need to develop their own data representations, nor will they have to deal with the full complexity of XML.
The developed tools are actively being used in our own projects, such as \MathML{\sc Ben}~\cite{Schubotz2018}, VMEXT~\cite{vmext17}, the Mathematical Language Processing (MLP) Project~\cite{disCicm14Mlp}, and 
HyPlag~\cite{DBLP:conf/iceis/GippMBPN14,Meuschke12,Meuschke2018a} for mathematical plagiarism detection.
Our hope is that the tools presented in this paper will help others to realize \MathML{} related projects in the future.

Our goal is to actively maintain and extend \name{}.
Plans for current and future work include enlarging the libraries of XPath and XQuery expressions and providing transformation tools for \MathML{} trees.
Furthermore, we are currently working on semantic enhancements of \MathML{} through the use of WikiData (as proposed in the \MathML{\sc Ben} project).
This semantic enhanced \MathML{} data can be used to calculate semantic distances through the calculation of distances between WikiData links.
Moreover, we will consider the development of convienient interfaces to other tools that deal with similar standars such as OpenMath. For instance         Py-OpenMath\footnote{\url{https://github.com/OpenMath/py-openmath}} is a project written in Python that allows one to parse XML formats and provides for conversion between Python objects to OpenMath representations.
Additionaly, we are considering to take advantage of tools for processing semantics on the document level. In particular MMT~\cite{DBLP:conf/mkm/Rabe13}, which is a framework for creation of a new language for partially formal mathematical knowledge.
We plan to maintain the current API for the foreseeable future.

%\vspace*{0.5em}
\noindent
{\bf Acknowledgements}
We would like to thank Felix Hamborg, Vincent Stange, Jimmy Li, Telmo Menezes, and Michael Kramer for contributing to the \name{} project.
We are also indebted to Akiko Aizawa for her advice and for hosting us as visiting researchers at the National Institute of Informatics (NII) in Tokyo.
This work was supported by the FITWeltweit program of the German Academic Exchange Service (DAAD) as well as the German Research Foundation (DFG) grant no. GI 1259/1.

%\vspace*{-1.5em}
\printbibliography
\end{document}